\documentstyle[twocolumn,aps,epsfig]{revtex}

\begin{document}
\title{Black Hole Relics in Large Extra Dimensions}

\author{Sabine Hossenfelder\thanks{email: hossi@th.physik.uni-frankfurt.de}, 
Marcus Bleicher, Stefan Hofmann, Horst St\"ocker}

\address{Institut f\"ur Theoretische Physik\\ J. W. Goethe-Universit\"at\\
Robert-Mayer-Str. 8-10\\ 60054 Frankfurt am Main, Germany}

\author{Ashutosh V. Kotwal}

\address{Duke University, Dept. of Physics\\Box 90305, Durham N.C. 27708-0305, USA}

\maketitle

\noindent
\begin{abstract}
Recent calculations applying statistical mechanics  indicate that in a
setting with  compactified large extra dimensions a black hole might evolve into a
(quasi-)stable state with  mass close to the new fundamental scale $M_f$. 
Black holes and therefore their relics might be produced at the {\sc LHC} in the case of 
extra-dimensional topologies.  In this energy regime, Hawking's evaporation
scenario is modified due to energy conservation and quantum effects.
We reanalyse the evaporation of small black holes including the quantisation 
of the  emitted radiation due to the finite surface of the black hole.
It is found that observable stable black hole relics  with masses $\sim 1-3 M_f$ would form which
could be identified by a delayed single jet with a corresponding hard momentum kick to the relic and
by ionisation, e.g. in a {\sc TPC}. 
\vspace{1cm}
\end{abstract}

The idea of Large eXtra Dimensions ({\sc LXD}s) which was recently 
proposed in \cite{add1,add2,add3,Randall:1999vf,Randall:1999ee} might allow to study interactions
at trans-planckian energies in the next generation collider experiments. 
Here, the hierarchy-problem is solved 
or at least reformulated in a geometric language by the existence of $d$ compactified {\sc LXD}s 
in which only the gravitons can propagate. The standard-model particles are bound to our 4-dimensional 
sub-manifold, often called our 3-brane. 

The strength of a force at a distance $r$ generated by a charge depends on the 
number of space-like dimensions. For distances smaller than the compactification length $L$, 
the gravitational interaction drops faster compared to the other interactions.
For distances much bigger than $L$ gravity is described by the well known potential 
law $\propto 1/r$. However, starting from $r\geq L$ the force lines are diluted into
the extra dimensions resulting in a smaller effective coupling constant for gravity.  

This scenario would lead to the following relation between the four-dimensional Planck mass $m_{\rm p}$ and 
the higher dimensional Planck mass $M_f$, which is the new fundamental scale of the theory
\begin{eqnarray} \label{master}
m_{\rm p}^2 = L^d M_f^{d+2} \quad.
\end{eqnarray}

The lowered fundamental scale would induce a vast number of 
observable phenomena for quantum gravity at 
energies in the range $M_f$. In fact, the non-observation of these predicted features gives 
first constraints on the parameters of the model, the number of extra dimensions $d$ and the fundamental scale 
$M_f$ \cite{Revcon,Cheung}. On the one hand, this scenario would have major 
consequences for cosmology and 
astrophysics  such as the modification of inflation in the early universe and enhanced 
supernova-cooling due to graviton emission \cite{add3,Cullen,Probes,astrocon,GOD}. 
On the other hand, additional processes had to be expected in high-energy collisions \cite{Rums}: production of real and 
virtual gravitons \cite{enloss1,enloss2,Hewett,Nussi,Rizzo} and the creation of black holes at 
energies that can be achieved at colliders in the near future.
 
Especially the possibility of black hole production in 
{\sc LXD}s at the {\sc LHC} and from cosmic rays has received great attention 
\cite{adm,ehm,Banks:1999gd,gid,dim,1,2,3,Feng:2001ib,Emparan:2001kf,anchor,Ringwald:2001vk,Uehara:2001yk,Landsberg:2001sj,Kotwal:2002wg}.
Black holes produced in such interactions would be tiny and may decay on fm/c time scales \cite{3}.
Thus, the decay of these objects (black holes, p-branes, string balls) could be studied in
detail in the laboratory. 
Unfortunately, very little is known about the  final stages of black hole evaporation. 
Extensive speculations about the final fate of black holes have been brought forward in the literature and 
will be discussed in detail later. 
The general notion is that a small black hole stops evaporating particles when its mass approaches 
the Planck scale, resulting in the exciting possibility of forming a (quasi-)stable relic.
In this letter, we study a model for the radiation from small black holes 
assuming a geometrical quantisation of the emitted radiation. The quantisation of the radiation
can lead to black hole relics of masses around 1-3 TeV with small electric charge. Those relics 
might be observable at the {\sc LHC}.

Let us start with the properties of black holes which may be accessible in the next generation colliders. A black hole which might be produced with $\sqrt{s}\approx 10$ TeV would 
have a radius much smaller than the size $L$ of the {\sc LXD}s \cite{3}. In this case one can neglect the periodic boundary conditions due to compactification and approximate the space-time to be spherically symmetric. For these black holes, the  metric is given by the $(d+4)$ - dimensional 
Schwarzschild metric \cite{my}.

Following Ref.\cite{my}  and implying the extra dimensions via Eq. (\ref{master}), the 
Schwarzschild radius is given by
\begin{equation} \label{ssr}
R_S^{d+1} = \frac{2}{d+1}\frac{M}{M_f^{d+2}} \quad. 
\end{equation}
The metric takes the familiar form 
\begin{equation} \label{ssmetric}
{\mathrm{d}} s^2 = -\gamma(r) {\mathrm{d}}t^2 + \frac{1}{\gamma(r)} {\mathrm{d}}r^2 + 
{\mathrm{d}} \Omega_{(d+3)}^2  
\end{equation}
with ${\mathrm{d}}\Omega_{(d+3)}$ being the surface element of the $(d+3)$-dimensional sphere, containing $(d+2)$ 
angles and $\gamma(r)$ given by
\begin{equation}
\gamma(r)=1 - \left( \frac{R_S}{r} \right) ^{d+1} \quad.
\end{equation}
From this one gets the surface gravity:
\begin{equation} \label{kappa}
\kappa = \frac{d+1}{2}\frac{1}{R_S}  \quad,
\end{equation}
which is the Newtonian force at the horizon
in the Schwarzschild case.
The surface of the black hole is 
\begin{equation}  
{\cal A} = \Omega_{(d+3)} R_S^{d+2}  \quad,
\end{equation}
 with $\Omega_{(d+3)}$ denoting the surface of the unit $d+3$-sphere
\begin{eqnarray}
\Omega_{(d+3)} = \frac{2 \pi^{\frac{d+3}{2}}}{\Gamma({\frac{d+3}{2}})}\;\;.
\end{eqnarray}

The production cross section for black holes in parton-parton or $\nu N$ collisions 
can be estimated on geometrical grounds and is of 
order $\sigma(M)\approx \pi R_S^2$ \cite{Banks:1999gd,Thorne:1972ji,Giddings:2001ih,Solodukhin:2002ui}.  
Detailed studies to support this estimate 
can be found in  the recent works by Jevecki and Thaler\cite{Thaler} and by Eardley and 
Giddings\cite{Giddings2}. Assuming the validity of the classical approximation and  
setting $M_f\sim 1$~TeV and $d=2$ one finds $\sigma \approx 1$~TeV$^{-2}\approx 400$~pb.
The luminosity of pp interactions at the {\sc LHC} would then allow the production of 
approximately $\approx 10^8$ black holes per year \cite{dim}. 

The fate of these small black holes is difficult to estimate.
There final evaporation is closely connected to the  information loss puzzle.
The black hole emits thermal radiation, whose sole property is the
temperature, regardless of the initial state of the collapsing matter. 
So, if the black hole completely decays into statistically
distributed particles, unitarity can be violated. This happens when 
the initial state is a pure quantum state and then evolves into a mixed state \cite{Hawk82,Preskill,Novi}.

When one tries to avoid the information loss problem two possibilities are left. The information
is regained by some unknown mechanism or a stable black hole remnant is formed which keeps the
information. Besides the fact that it is unclear in which way the information should escape the horizon 
\cite{escape1,escape2,escape3,escape4,escape5,escape6} there are several other arguments for 
black hole relics \cite{relics1,relics2,relics3,relics4}:
\begin{itemize}
\item
The uncertainty relation: The Schwarzschild radius of a black hole with Planck mass 
is of the order  of the Planck length. Since the Planck length is the wavelength corresponding to a particle of
Planck mass, a problem arises when the mass of the black hole drops below Planck mass. 
Then one has trapped a higher mass, $M\geq M_f$, inside a volume which is smaller than allowed by the uncertainty 
principle \cite{39}. To avoid this problem, Zel'dovich has proposed that black holes with masses below 
Planck mass should be associated with stable elementary particles \cite{40} 

\item
Corrections to the Lagrangian: The introduction of additional terms, which are quadratic in the curvature, yields a dropping of the evaporation temperature towards zero \cite{Barrow,Whitt}. This holds also for 
extra dimensional scenarios \cite{my2} and is supported  by calculations in the low energy limit 
of string theory \cite{Callan,Stringy}.

\item
Further reasons for the existence of relics have been suggested to be black holes with axionic charge
\cite{axionic}, the modification of the Hawking temperature due to quantum hair \cite{hair} or magnetic 
monopoles \cite{magn}. Coupling of a dilaton field to gravity also yields relics, with detailed 
features depending on the dimension of space-time \cite{dilaton1,dilaton2}.

\end{itemize}
 
Let us now compare the classical micro-canonical emission scenario to an approach that
takes into account the effects of the geometrical quantisation of the emitted radiation.
Note that this approach is different from a quantisation of the event horizon. 
In the present model, all horizon configurations can still be realized. 

Generally,  black holes emit particles via the Hawking mechanism \cite{Hawk1,Hawk2}. 
The temperature of the radiation is:
\begin{equation} \label{tempkappa}
T = \frac{\kappa}{2 \pi} \quad,
\end{equation}
with $\kappa$ given by (\ref{kappa}). 
From dimensional aspects we expect the entropy to be $S \propto {\cal{A}} M_f^{d+2}$. 
From Thermodynamics we know that,
\begin{eqnarray}
\frac{\partial S}{\partial M} = \frac{1}{T} \;\;,
\end{eqnarray}
where $M$ is interpreted  as the conserved energy of the system.

Inserting Eqs. (\ref{ssr}), (\ref{kappa}) and (\ref{tempkappa}) one obtains a mass dependence of $1/T$ 
with the exponent $(1/(d+1))$. Integration yields
\begin{equation} \label{entropied}
S(M) = \frac{d+1}{d+2} \frac{M}{T} = 2 \pi \frac{d+1}{d+2}  \left( M_f R_S \right)^{d+2} \quad.
\end{equation}
In the limit where the energies of the emitted particles are  small 
compared to the mass of the black hole the grand-canonical ensemble can be used.
The number density of particles with energy $\omega$ is then
\begin{equation} \label{nkan}
n(\omega) = 
\frac{1}{\exp{\frac{\omega}{T}}-1} \;\;.
\end{equation}
with which one derives the higher dimensional analogue of the Stefan-Boltzmann law \cite{Agnese,Harms1}
\begin{eqnarray} \label{epsil}
\varepsilon &=&   \frac{\Omega_{(d+3)}^2}{(2 \pi)^{d+3}} \Gamma(d+4) \zeta(d+4) T^{d+4} \;\;.
\end{eqnarray} 
The spectrum of the emitted radiation has a maximum at frequencies of the order of the temperature. 
Thus, when the mass of the black hole approaches the Planck scale, the energy of the emitted 
quanta can no longer be neglected. Since we are interested in the late evaporation stage, 
when the black hole is small and hot, an appropriate statistical description
is given by  the micro-canonical ensemble \cite{Harms1,Harms2,Harms3}. 
Here, the single-particle number density is given by
\begin{equation} \label{nmik}
n(\omega) =  \frac{\exp{[S(M-\omega)]}}{{\rm exp}[S(M)]}\quad,
\end{equation}
where $S$ denotes the entropy of the black hole. 

Let us shortly examine the limit of huge black hole masses 
in the micro-canonical approach.
The limit of the grand-canonical number density is 
Wien's limit, i.e. $n(\omega)=\exp(-\omega/T)$. 
In the micro-canonical case
\begin{equation}
\ln n(\omega) \approx - 2 \pi (M_f R_S)^{d+2}   \frac{\omega}{M} 
= - \frac{4 \pi R_S}{d+1} \omega = - \frac{\omega}{T} 
\end{equation}
in leading order in $\omega/M$. Thus, Wien's limit is recovered.

Next we investigate the multi-particle spectrum in the micro canonical description:
\begin{equation} \label{mehrteil}
n(\omega) = 
\sum_{j=1}^{\lfloor \frac{M}{\omega} \rfloor} \frac{{\rm exp}[S(M-j\omega)]}{{\rm exp}[S(M)]} \quad,
\end{equation}
with $\lfloor x \rfloor$ being the smallest integer next to $x$. 
This cut-off assures that the total energy of the emitted quantum  
does not exceed the mass of the black hole.
After substituting $x=M-j\omega$ one obtains for the total energy density which is 
radiated off by the black hole
\begin{eqnarray}
\varepsilon &=& \frac{\Omega_{(d+3)}}{(2 \pi)^{3+d}} 
{\mathrm e}^{-S(M)} \sum_{j=1}^{\infty} \frac{1}{j^{d+4}} \times \nonumber \\
&&\int_0^{M} {\mathrm e}^{S(x)} (M-x)^{3+d} {\mathrm d}x \quad.
\end{eqnarray}
The evaporation rate per degree of freedom is given by
\begin{eqnarray}  
\frac{{\mathrm d}M}{{\mathrm d}t} &=& \frac{\Omega_{(d+3)}^2}{(2\pi)^{d+3}} R_S^{2+d} \zeta(d+4) \; 
{\rm e}^{-S(M)} \times \nonumber \\
&&\int_{0}^{M} {\rm e}^{S(x)} (M-x)^{(3+d)} {\mathrm d}x \quad.\label{mdoteq}
\end{eqnarray}

\begin{figure}[h]
\vskip 0mm
\vspace{-1.5cm}
\centerline{\psfig{figure=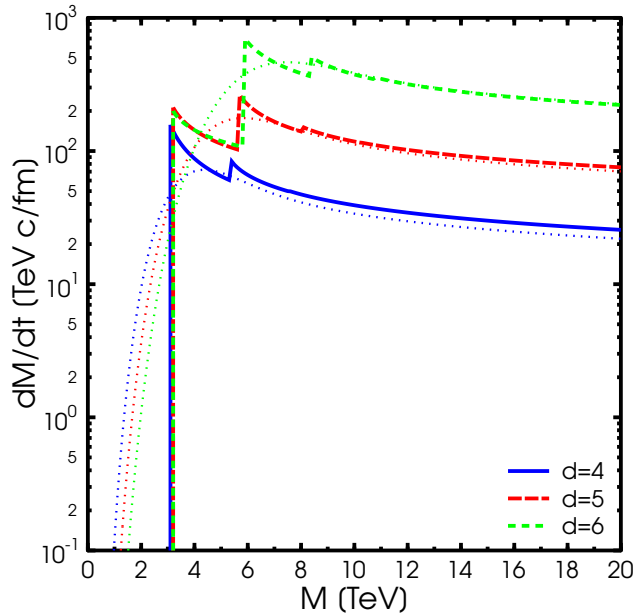,width=4.0in}}
\vskip 2mm
\caption{Black hole evaporation rate per degree of freedom 
as a function of mass, from top to bottom for $d=6,5,4$. 
The thin dotted lines show the standard micro-canonical scenario.
The thick lines denote the calculations with the discrete energy spectrum.  
\label{dmdtq}}
\end{figure}

Fig. \ref{dmdtq} (thin dotted lines) shows the 
evaporation rate (\ref{mdoteq}) as a function of the initial mass $M$ of the 
black hole.  In the limit $M\rightarrow \infty$, the micro canonical evaporation rate 
reproduces the Hawking evaporation in $(d+3)$ space-like dimensions:
\begin{eqnarray}
\lim_{M\gg M_f} \frac{{\rm d}M}{{\rm d}t} &=& {\cal{A}}_D \frac{\Omega_{(d+3)}}{(2 \pi)^{d+3}} 
\int_0^{\infty} \; \frac{\omega^{3+d}\; {\rm d} \omega}{\exp{(\omega T^{-1})} -1}\nonumber \\
&=& \frac{\Omega(d+4)^2}{(2 \pi)^{d+3}} \Gamma(d+4) \zeta(d+4) \; R_S^{2+d}  T^{d+4}  \quad, \nonumber
\end{eqnarray} 
which is, using (\ref{tempkappa}),
 \begin{eqnarray}
\lim_{M\gg M_f} \frac{{\rm d}M}{{\rm d}t} &=& \frac{\Omega_{(d+3)}^2}{(2 \pi)^{2d+7}} 
\left(\frac{d+1}{2}\right)^{d+4} \Gamma(d+4) \frac{\zeta(d+4)}{ R_S^{2}} \nonumber \\
&\propto& M_f^2 \left( \frac{M}{M_f}\right)^{-\frac{2}{d+1}} \quad. \nonumber
 \end{eqnarray}

Now we address the model with geometrical quantisation of the emitted radiation. The radiation 
of a black hole derived by  semi-classical quantum field
theory in curved space \cite{BiDa} yields a black body spectrum in $d+3$ dimensions. Spherical symmetry 
is taken into account by making the usual separation ansatz for the wave equation:
\begin{eqnarray}
\psi(r,\Omega) = \frac{1}{r^{d+1}} Y_{lm}(\Omega) \phi(r) \quad,
\end{eqnarray}

which factorizes the full wave function into an amplitude $1/r^{d+1}$, a radial wave 
function $\phi$ and a set of spherical harmonics $Y_{lm}$.
Here $\Omega$ is an abbreviation for the occurring $d+2$ angles. The
coefficients in the multipole expansion suppress the energy emission in higher
order harmonics, therefore we consider only the $l=0$ mode.  
In this case the dispersion relation is the usual one for massless particles $k^2=\omega^2$. 
The boundary conditions 
of the black body lead to a restriction of the possible momenta \footnote{%
Note that the energy levels of the radiation in the monopole contribution are similar to the 
energy levels of a cubic black body radiator of side length $2 R_S$ that was discussed in Ref. \cite{stehof}.
}.

This results in a geometrical quantisation of the momentum spectrum  
\begin{eqnarray}
k_l = \frac{\pi l}{R_S} \quad.
\end{eqnarray}
As a consequence, the
multi-particle spectrum (\ref{mehrteil}) is modified, since
the emitted particles have energies in integer multiples of a minimal 
energy quantum $\Delta \omega = \pi / R_S$:

\begin{equation} \label{mehrteilq}
n(l) = 
\sum_{j=1}^{\lfloor \frac{M}{l \Delta \omega} \rfloor} \frac{{\rm exp}[S(M-j l \Delta \omega)]}{{\rm exp}[S(M)]} 
\Theta\left(M - l \Delta \omega \right) \quad.
\end{equation}
Here the $\Theta$-function cuts off the spectrum when the energy of one particle 
exceeds the mass of the black hole. The energy density of the radiation is derived by  summation over 
momentum space
\begin{eqnarray} \label{epsq}
\varepsilon &=&  \frac{\Omega_{(d+3)}}{(2 \pi)^{3+d}}  
\Delta \omega  
\sum_{l=1}^{ \lfloor \frac{M}{\Delta \omega} \rfloor } n(l)  
\left( l \Delta \omega \right)^{d+3}  \quad.
\end{eqnarray}

In the limit of large black hole masses $M \gg M_f$ one has $\Delta \omega  \to 0$ 
and regains the continuous emission spectrum from Eq. (\ref{epsq}).   
The evaporation rate with respect to the geometrically quantised spectrum is 
shown in Fig. \ref{dmdtq} (thick lines) and compared to the continuous spectrum case (thin dotted lines).  

The spacing of energy levels gets smaller with increasing $M$, and whenever it is
possible to occupy an additional level the evaporation rate exhibits a step. 
These steps naturally appear in spacings $\approx \pi$~TeV, because in 
the mass range of interest it is $R_S \approx 1/M_f$ -- the exact value thereby depending 
on the number of extra dimensions.
 
The evaporation process occurs in quantised steps. It can not proceed further 
when the lowest lying quantum state allowed exceeds already the mass of the 
black hole. Evaporation is halted at a finite mass value for a certain fraction of initial masses
above the fundamental scale. It should be noted that the transverse momentum spectra
of the radiation is modified as compared with a simple extrapolation of the Hawking formula 
to small masses:
Here the final quanta possess energies of the order $1/R_S$, and not a continuous spectrum 
with $T(M\approx 1~{\rm TeV})$.

If the mass were slightly above the Planck scale, it would  still be possible for the black hole to 
emit a particle carrying away most of its energy. This might leave a stable relic with 
mass below Planck mass. 
However, since physics below the Planck scale is unknown, 
we cannot be sure about the existence and properties 
of relics with masses below the fundamental scale.
Thus, in the following we will focus on those relics with masses above the fundamental scale.   
 
\begin{figure}[h]
\vskip 0mm
\vspace*{-1.0cm}
\centerline{\psfig{figure=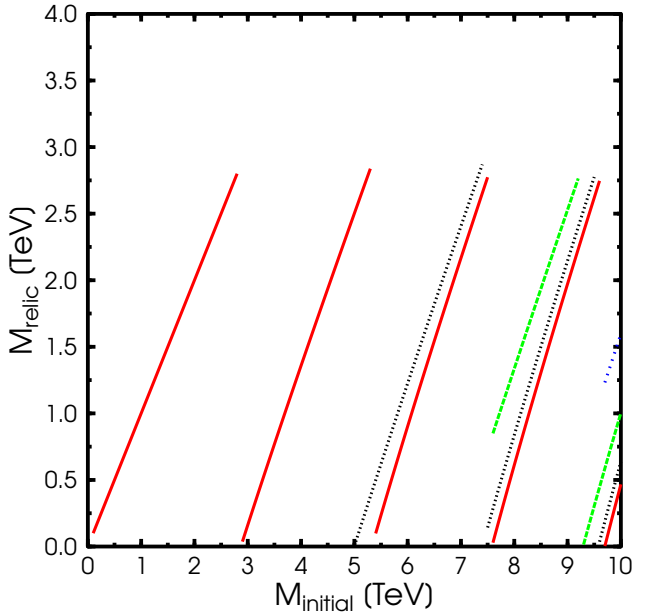,width=4.0in}}
\vskip 2mm
\caption{Possible final relic masses ($M_{\rm relic}$) after the evaporation process from a black hole
of initial mass $M_{\rm initial}$. The calculation is for $d=3$ extra dimensions and $M_f=1$~TeV. 
The solid lines belong to the most probable case, other lines include the emission of higher modes.   
\label{miniend}}
\end{figure}

It is interesting to ask for the spectrum of final relic 
masses $M_{\rm relic}$
depending on the initial mass $M_{\rm initial}$
of the black hole.
This relation is shown in Fig. \ref{miniend} for the case $d=3$. The most
probable case  is the exclusive emission of minimal energy quanta 
during the evaporation process, depicted by solid lines in Fig. \ref{miniend}.
Black holes with $M_{\rm initial} \leq 3 M_f$ are stable 
with $M_{\rm relic}=M_{\rm initial}$.
The inclusion of higher modes in the evaporation process,
which becomes more important with increasing initial mass, is shown 
by the dashed and dotted lines.
When going even further, to masses $M_{\rm initial} \gg M_f$, the whole range of end 
masses would become accessible. Because the black holes accessible at the {\sc LHC} would have 
mainly masses slightly above $M_f$, most 
of these black holes were stable with masses around $M_f$.  
 
Indirect constraints on the existence of black hole relics can be obtained from
the decay of primordial black holes. Their modified and dimension dependent 
energy spectrum influences observables, e.g. the cosmic microwave background and 
the baryogenesis \cite{Griff,Liddle}.   
Furthermore, the black hole relics from primordial density fluctuations may be 
a candidate for dark matter \cite{Blais,Barrow}. Upper limits on the relative contribution 
of those relics to the critical energy density in the universe are on the order 
of $\Omega_{\rm relics}=0.1-1$ \cite{macgibbon}.
Thus, the observation of relics in a collider experiment is of highest interest.

The final and most interesting question is: How could one observe these relics?
\begin{itemize}
\item
A black hole relic with a mass of $\approx 3$ TeV  would have a spectrum 
that just fails to allow for a last
emission of a quantum (cf. Fig. \ref{miniend}). Therefore, if its mass were increased only 
slightly by the energy $\Delta E$,   
it would enable the black hole to evaporate again, emitting a high energetic quantum and leaving 
a tiny mass relic. This might result in a delayed flash of hard photons, leptons 
or {\sc QCD} jets compared to the collision dynamics encountered at the {\sc LHC}.   
The fraction of  black holes evaporating in this manner 
can be estimated from the mass spectrum of black holes produced at the {\sc LHC} 
and is $\Delta E/100~{\rm TeV}^{-1} \sim 10^{-4}-10^{-5}$ for $\Delta E= 1$~GeV.
Note that relics from primordial black holes  might also lead to observable air showers, 
if a black hole relic  evaporates in the atmosphere.

\item
A certain fraction of the black holes produced in parton-parton collisions 
would carry a small charge of 
order $e$. This might allow to identify the
charged black hole relics, e.g.  by ionisation in a time projection chamber.

\item
The thermal evaporation spectrum would be much softer as expected in the literature \cite{dim} which
assumes total decay of TeV black holes. However, the final stages would be governed by non-thermal
particle emission.
\end{itemize}

To summarise, we have analysed the late stages of black hole evaporation, including the 
geometrical quantisation of the emitted radiation.
In this  model setting, the production of stable black hole relics would be possible at
the {\sc LHC}. These relics may be observable in a late burst of jets if they capture additional energy.
Charged black hole relics may eventually be directly detected, e.g. in a {\sc TPC} by ionisation.

\section*{Acknowledgements}
The authors acknowledge fruitful discussions with L.~ Gerland and A.~ Dumitru. S.~Hossenfelder wants to thank the Land Hessen for financial support.


\begin{thebibliography}{99}

\bibitem{add1} N.~Arkani-Hamed, S.~Dimopoulos  and  G.~Dvali, 
Phys. Lett. B {\bf 429}, 263-272 (1998)
{ [arXiv:hep-ph/9803315]} 

\bibitem{add2} I.~Antoniadis, N.~Arkani-Hamed, S.~Dimopoulos  and  G.~Dvali, 
Phys. Lett. B {\bf 436}, 257-263 (1998) { [arXiv:hep-ph/9804398]} 

\bibitem{add3} N.~Arkani-Hamed, S.~Dimopoulos  and  G.~Dvali, 
Phys. Rev. D {\bf 59}, 086004 (1999) { [arXiv:hep-ph/9807344]}.

\bibitem{Randall:1999vf}
L.~Randall and R.~Sundrum,
Phys.\ Rev.\ Lett.\  {\bf 83} (1999) 4690
[arXiv:hep-th/9906064].

\bibitem{Randall:1999ee}
L.~Randall and R.~Sundrum,
Phys.\ Rev.\ Lett.\  {\bf 83} (1999) 3370
[arXiv:hep-ph/9905221].

\bibitem{Revcon} Y.~Uehara, 
Mod.\ Phys.\ Lett.\ A {\bf 17} (2002) 1551
[arXiv:hep-ph/0203244].

\bibitem{Cheung} K.~Cheung, 
{ [arXiv:hep-ph/0003306]}.
Presented at 7th International Symposium on Particles, Strings and Cosmology
(PASCOS 99), Granlibakken, Tahoe City, California, 10-16 Dec 1999.
Published in *Lake Tahoe 1999, Particles, strings and cosmology* 145-155

\bibitem{Cullen} S.~Cullen  and  M.~Perelstein, 
Phys. Rev. Lett. {\bf 83}, 268 (1999) { [arXiv:hep-ph/9903422]}.

\bibitem{Probes} 

J.~Hewett and M.~Spiropulu,
Ann.\ Rev.\ Nucl.\ Part.\ Sci.\  {\bf 52} (2002) 397
[arXiv:hep-ph/0205106].


\bibitem{astrocon} V.~Barger, T.~Han, C.~Kao  and  R.-J.~Zhang, 
Phys. Lett. {\bf B 461}, 34 (1999), { [arXiv:hep-ph/9905474]}.

\bibitem{GOD} C.~Hanhart, J.~A.~Pons, D.~R.~Phillips  and  S.~Reddy, 
Phys. Lett. {\bf B 509} 1-9 (2001), { [astro-ph/0102063]}.


 
\bibitem{Rums}  E.~A.~Mirabelli, M.~Perelstein  and  M.~E.~Peskin, 
Phys. Rev. Lett. {\bf 82} 2236-2239 (1999) { [arXiv:hep-ph/9811337]}

\bibitem{enloss1} G.~F.~Giudice, R.~Rattazzi  and  J.~D.~Wells, 
Nucl.Phys. B544 3-38 (1999) { [arXiv:hep-ph/9811291]}.

\bibitem{enloss2} G.~F.~Giudice, R.~Rattazzi  and  J.~D.~Wells, 
 Nucl.Phys. {\bf B595} 250-276 (2001) { [arXiv:hep-ph/0002178]}  

\bibitem{Hewett} J.~L.~Hewett, 
Phys.Rev.Lett. 82 4765-4768 (1999) { [arXiv:hep-ph/9811356]}.

\bibitem{Nussi}   S.~Nussinov  and  R.~Shrock, 
Phys.Rev. D59 105002 (1999) { [arXiv:hep-ph/9811323]}.
  

\bibitem{Rizzo} T.~G.~Rizzo, 
{ [arXiv:hep-ph/9910255]}.
Proceedings of 2nd International Conference Physics Beyond the Standard
Model: Beyond the Desert 99: Accelerator, Nonaccelerator and Space
Approaches, Ringberg Castle, Tegernsee, Germany, 6-12 Jun 1999.
Published in *Tegernsee 1999, Beyond the desert 1999* 23-44 

\bibitem{adm}
P.C.~Argyres, S.~Dimopoulos, and J.~March-Russell,
Phys. Lett. {\bf B441}, 96 (1998)
{ [arXiv:hep-th/9808138]}.


\bibitem{ehm}
R.~Emparan, G.T.~Horowitz, and R.C.~Myers,
Phys. Rev. Lett. {\bf 85}, 499 (2000) { [arXiv:hep-th/0003118]}.

\bibitem{Banks:1999gd}
T.~Banks and W.~Fischler, 
 { [arXiv:hep-th/9906038]}.

\bibitem{gid}
S.~B.~Giddings and S.~Thomas, 
Phys.\ Rev.\ D {\bf 65}, 056010 (2002) { [arXiv:hep-ph/0106219]}.


\bibitem{dim} S.~Dimopoulos  and  G.~Landsberg,  
Phys. Rev. Lett. {\bf 87}, 161602 (2001) { [arXiv:hep-ph/0106295]}.

\bibitem{1}
S.~Hofmann, M.~Bleicher, L.~Gerland, S.~Hossenfelder, K.~Paech and H.~St\"ocker,
J.\ Phys.\ G {\bf 28} (2002) 1657.

\bibitem{2} M.~Bleicher, S.~Hofmann, S.~Hossenfelder  and  H.~St\"ocker, 
Phys. Lett. {\bf B548}, 73-76 (2002), { [arXiv:hep-ph/0112186]}.


\bibitem{3} 
S.~Hossenfelder, S.~Hofmann, M.~Bleicher and H.~St\"ocker,
Phys.\ Rev.\ D {\bf 66} (2002) 101502
[arXiv:hep-ph/0109085].

 
\bibitem{Feng:2001ib}
J.~L.~Feng and A.~D.~Shapere, 
Phys.\ Rev.\ Lett.\  {\bf 88}, 021303 (2002) 
{ [arXiv:hep-ph/0109106]}.

 
\bibitem{Emparan:2001kf}
R.~Emparan, M.~Masip and R.~Rattazzi,
Phys.\ Rev.\ D {\bf 65},  064023  (2002) { [arXiv:hep-ph/0109287]}.
 
\bibitem{anchor}
L.~A.~Anchordoqui, J.~L.~Feng, H.~Goldberg and A.~D.~Shapere,
Phys.\ Rev.\ D {\bf 65} (2002) 124027
[arXiv:hep-ph/0112247].

  
\bibitem{Ringwald:2001vk}
A.~Ringwald and H.~Tu, 
Phys.\ Lett.\ B {\bf 525}, 135 (2002) { [arXiv:hep-ph/0111042]}.

\bibitem{Uehara:2001yk}
Y.~Uehara, 
Prog.\ Theor.\ Phys.\  {\bf 107},  621 (2002) { [arXiv:hep-ph/0110382]}.

\bibitem{Landsberg:2001sj}
G.~Landsberg, 
Phys.\ Rev.\ Lett.\  {\bf 88}, 181801 (2002) { [arXiv:hep-ph/0112061]}.

\bibitem{Kotwal:2002wg}
A.~V.~Kotwal and C.~Hays,
Phys.\ Rev.\ D {\bf 66} (2002) 116005
[arXiv:hep-ph/0206055].

\bibitem{my} R.~C.~Myers  and  M.~J.~Perry 
Ann. Phys. {\bf 172}, 304-347 (1986)

\bibitem{Thorne:1972ji}
K.~S. Thorne, Nonspherical gravitational collapse: A short review, in J R Klauder, Magic Without
Magic, San Francisco 1972, 231-258.


\bibitem{Giddings:2001ih}
S.~B.~Giddings,
in {\it Proc. of the APS/DPF/DPB Summer Study on the Future of Particle Physics (Snowmass 2001) } ed. N.~Graf,
eConf {\bf C010630} (2001) P328
[arXiv:hep-ph/0110127].
 

\bibitem{Solodukhin:2002ui}
S.~N.~Solodukhin,
Phys.Lett. {\bf B 533}, 153-161, (2002),  
{ [arXiv:hep-ph/0201248]}.

\bibitem{Thaler}
A.~Jevicki and J.~Thaler, 
 Phys.Rev. {\bf D 66}, 024041 (2002),  { [arXiv:hep-th/0203172]}. 

\bibitem{Giddings2}
D.~ M.~Eardley  and  S.~B.~Giddings, 
Phys. Rev. {\bf D 66}, 044011 (2002), { [arXiv:gr-qc/0201034]}.

 

 
\bibitem{Hawk82} S.~Hawking, 
Commun. Math. Phys. {\bf 87},395 (1982). 

 

\bibitem{Preskill} J.~Preskill, 
{ [arXiv:hep-th/9209058]}
Presented at International Symposium on Black holes, Membranes, Wormholes
and Superstrings, Woodlands, TX, 16-18 Jan 1992.
In *Houston 1992, Proceedings, Black holes, membranes, wormholes and
superstrings* 22-39.
 
\bibitem{Novi} I.~D.~Novikov  and  V.~P.~Frolov, ''{\sl{Black Hole Physics}}'', 
Kluver Academic Publishers (1998)

  

\bibitem{escape1} D.~N.~Page, Phys. Rev. Lett. {\bf 44}, 301 (1980).

\bibitem{escape2} G.~t'Hooft, Nucl. Phys. {\bf B 256} 727 (1985).

\bibitem{escape3} A.~Mikovic, Phys. Lett. {\bf 304 B}, 70 (1992).

\bibitem{escape4} E.~Verlinde  and  H.~Verlinde, Nucl. Phys. {\bf B 406}, 43 (1993).

\bibitem{escape5} L.~Susskind, L.~Thorlacius  and  J.~Uglum, Phys. Rev. {\bf D 48}, 3743 (1993).

\bibitem{escape6} D.~N.~Page, Phys. Rev. Lett. {\bf 71} 3743 (1993).

\bibitem{relics1} Y.~Aharonov, A.~Casher  and  S.~Nussinov,
Phys. Lett. {\bf 191 B}, 51 (1987).

\bibitem{relics2} T.~Banks, \nolinebreak A.~Dab\nolinebreak holkar, \nolinebreak M.~R.~Dou\nolinebreak glas  and   M.~O'Loughlin,
Phys. Rev. {\bf D 45} 3607 (1992) 
{ [arXiv:hep-th/9201061]}.

\bibitem{relics3} T.~Banks  and  M.~O'Loughlin, 
Phys. Rev. {\bf D 47}, 540 (1993) { [arXiv:hep-th/9206055]}

\bibitem{relics4} T.~Banks, M.~O'Loughlin  and  A.~Strominger, 
Phys. Rev. {\bf D 47}, 4476 (1993) 
{ [arXiv:hep-th/9211030]}.

\bibitem{39} M.~A.~Markov, in: ''{\sl Proc. 2nd Seminar in Quantum Gravity}'', edited by M.~A.~Markov 
 and  P.~C.~West, Plenum, New York (1984).

\bibitem{40} Y.~B.~Zel'dovich, in: ''{\sl Proc. 2nd Seminar in Quantum Gravity}'', edited by M.~A.~Markov 
 and  P.~C.~West, Plenum, New York (1984).

\bibitem{Barrow} J.~D.~Barrow, E.~J.~Copeland  and  A.~R.~Liddle, 
Phys. Rev.  {\bf D 46}, 645 (1992).

\bibitem{Whitt} B.~Whitt, Phys. Rev. {\bf D 38}, 3000 (1988).

\bibitem{my2} R.~C.~Myers  and  J.~Z.~Simon, Phys. Rev. {\bf D 38}, 2434 (1988).

\bibitem{Callan} C.~G.~Callan, R.~C.~Myers  and  M.~J.~Perry, Nucl. Phys. {\bf B 311}, 673 (1988).

\bibitem{Stringy} S.~Alexeyev, A.~Barrau, G.~Boudoul, O.~Khovanskaya  and  M. Sazhin, 
Class. Quant. Grav. {\bf 19}, 4431-4444 (2002), { [arXiv:gr-qc/0201069]}

\bibitem{axionic} M.~J.~Bowick, S.~B.~Giddings, J.~A.~Harvey, G.~T.~Horowitz  and  A.~Strominger, 
Phys. Rev. Lett. {\bf 61} 2823 (1988).

\bibitem{hair} S.~Coleman, J.~Preskill   and  F.~Wilczek, 
Mod. Phys. Lett. {\bf A6} 1631 (1991). 

\bibitem{magn} K-Y.~Lee, E.~D.~Nair  and  E.~Weinberg, 
Phys. Rev. Lett. {\bf 68} 1100 (1992) { [arXiv:hep-th/9111045]}.

\bibitem{dilaton1} G.~W.~Gibbons  and  K.~Maeda, 
Nucl. Phys. {\bf B 298} 741 (1988).

\bibitem{dilaton2} T.~Torii  and  K.~Maeda, 
Phys. Rev. {\bf D 48} 1643 (1993).

\bibitem{Hawk1} S.~W.~Hawking, 
Comm. Math. Phys. {\bf 43}, 199-220 (1975).
 
\bibitem{Hawk2} S.~W.~Hawking, 
Phys. Rev. {\bf D 14}, 2460-2473 (1976).

\bibitem{Agnese} A.~G.~Agnese, M.~La Camera, E.~Recami, 
Il Nuovo Cimento 114B, 1367 (1999) { [arXiv:physics/9907008]}.

\bibitem{Harms1} R.~Casadio \& B.~Harms, 
Phys.Rev. {\bf D 64}, 024016 (2001), { [arXiv:hep-th/0101154]}.

\bibitem{Harms2} R.~Casadio \& B.~Harms, 
Phys.Lett. {\bf B 487}, 209-214 (2000),  { [arXiv:hep-th/0004004]}

\bibitem{Harms3} R.~Casadio \& B.~Harms, 
Int. J. Mod. Phys. {\bf A17}, 4635-4646 (2002), { [arXiv:hep-th/0110255]}

\bibitem{BiDa} N.~D.~Birell \& P.~C.~W.~Davies, ''{\sl{Quantum Fields in Curved
 Space}}'', Cambridge University Press (Cambridge, 1982).

\bibitem{stehof} A.~V.~Kotwal  and  S.~Hofmann, 
{ [arXiv:hep-ph/0204117]}.

 
\bibitem{Griff}
L.~M.~Griffiths, D.~Barbosa and A.~R.~Liddle,
{ [arXiv:astro-ph/9812125]}.


\bibitem{Liddle}
A.~R.~Liddle and A.~M.~Green,
Phys.\ Rept.\  {\bf 307} (1998) 125
{ [arXiv:gr-qc/9804034]}.

\bibitem{Blais}
D.~Blais, C.~Kiefer and D.~Polarski,
Phys.\ Lett.\ B {\bf 535} (2002) 11
{ [arXiv:astro-ph/0203520]}.

\bibitem{Barrow}
J.~D.~Barrow, E.~J.~Copeland and A.~R.~Liddle,
Phys.\ Rev.\ D {\bf 46} (1992) 645.


\bibitem{macgibbon}
J. H. MacGibbon, Nature {\bf 329}, 308 (1987)

\end{thebibliography}
\end{document}